\documentclass[aps,groupedaddress,nofootinbib]{revtex4}

\usepackage{graphicx}
\newcommand{\be}{\begin{equation}}
\newcommand{\ee}{\end{equation}}
\newcommand{\bea}{\begin{eqnarray}}
\newcommand{\eea}{\end{eqnarray}}	
\newcommand{\nn}{\nonumber\\}
\newcommand{\ba}{\begin{array}}
\newcommand{\ea}{\end{array}}

\newcommand{\w}{\omega}
\newcommand{\we}{\omega(\eta)}
\newcommand{\intw}{\int{\we d\e}}
\newcommand{\C}{\mathcal C}
\newcommand{\Ce}{\mathcal{C}(\eta)}

\newcommand{\Cp}{\mathcal{C}'}
\newcommand{\rfv}{\mid 0 \rangle_f}
\newcommand{\lfv}{_f\langle 0 \mid}
\newcommand{\rmv}{\mid 0 \rangle}
\newcommand{\lmv}{\langle 0 \mid}
\newcommand{\bpsi}{\bar{\psi}}
\newcommand{\vp}{\vec{p}}

\newcommand{\vq}{\vec{q}}
\newcommand{\vx}{\vec{x}}
\newcommand{\e}{\eta}

\newcommand{\lag}{\mathcal L}

\newcommand{\st}{\sin \theta}
\newcommand{\sst}{\sin^2 \theta}
\newcommand{\ssst}{\sin^3 \theta}
\newcommand{\ct}{\cos \theta}

\begin{document}

\title{Flavour Condensates in Brane Models and Dark Energy}


\author{Nick E. Mavromatos, Sarben Sarkar and Walter Tarantino}
%\thanks{}
\affiliation{King's College London, Department of Physics, Strand, London WC2R 2LS, UK.}


%\date{\today}

\begin{abstract}
In the context of a microscopic model of string-inspired foam, in which foamy structures are provided by brany point-like defects (D-particles) in space-time, we discuss flavour mixing as a result of flavour non-preserving interactions of (low-energy) fermionic stringy matter excitations with the defects. Such interactions involve splitting and capture of the matter string state by the defect, and subsequent re-emission. As a result of charge conservation, only electrically neutral matter can interact with the D-particles.
Quantum fluctuations of the D-particles induce a non-trivial space-time background; in some circumstances this could be akin to a cosmological Friedman-Robertson Walker expanding-Universe, with weak (but non-zero) particle production. Furthermore the D-particle medium can induce an MSW type effect. We have argued previously, in the context of bosons,  that the so-called flavour vacuum is the appropriate state to be used, at least for low-energy excitations, with energies/momenta up to a dynamically determined cutoff scale. Given the intriguing  mass scale provided by neutrino flavour mass differences from the point of view of dark energy, we evaluate the flavour-vacuum expectation value (condensate) of the stress-energy tensor of the $1/2$-spin fields with mixing in an effective low-energy Quantum Field Theory in this foam-induced curved space-time. We demonstrate, at late epochs of the Universe, that the fermionic vacuum condensate behaves as a fluid with negative pressure and positive energy; however the equation of state has $w_{\rm fermion} > -1/3$ and so the contribution of the fermion-fluid flavour vacuum  alone could not yield accelerating Universes. Such contributions to the vacuum energy should be considered as (algebraically) additive to the flavoured boson contributions, evaluated in our previous works; this should be considered as natural from (broken) target-space supersymmetry that characterises realistic superstring/supermembrane models of space-time foam. The boson fluid is also characterised by positive energy and negative pressure, but its equation of state is, for late eras, close to $w_{\rm boson} \to -1$, and hence overall the D-foam universe appears accelerating at late eras.
\end{abstract}

\maketitle

\section{Introduction \label{sec1}}

During recent years it has been suggested that a certain, mathematically consistent, treatment of flavour mixing in Quantum Field Theory could have implications at a cosmological scale \cite{dark energy}. Specifically, adopting a Fock-space quantization formalism for the ``flavour'' states~\cite{vitiello,ji}, one can define a new vacuum state which, in the thermodynamic limit, is orthogonal to the mass-eigenstate vacuum. This orthogonality, in fact,
extends to the entire set of Fock-space states constructed out of the flavour vacuum, relative to those constructed out of the mass-eigenstate vacuum. It has been claimed~\cite{giusti} that the flavour-vacuum formalism, although mathematically consistent, nevertheless leads to no physically different predictions from the conventional formalism.
However, the authors of \cite{henning} have argued that probability conservation is only realised within the flavour state vacuum in quantum field theories with mixing; moreover the oscillation probability among flavours is modified, compared to the traditional formalism, by extra terms which, although small, nevertheless are in principle experimentally detectable. In this sense they postulated that in such cases the flavour vacuum is the physical vacuum.

It has then been demonstrated~\cite{dark energy} that the vacuum condensate due to fermion particle mixing, evaluated in the (physical) flavour vacuum, seems to behave as a source of Dark Energy, in the sense of yielding a non-trivial flavour-vacuum energy density. In a series of papers~\cite{conundrum}, following the initial work of \cite{dark energy}, it has also been argued that the fluid of flavour fermions behaves as an ideal one, in the cosmological sense, with a simple equation of state, which depends on the
Universe's epoch.

However, these calculations have been performed in the context of a Minkowski space-time Quantum Field Theory, despite the fact that flavour mixing gives rise to a non-trivial space-time cosmological background. A consistent treatment, therefore, requires consideration of the Fock-space vacuum in the presence of such
cosmological space-times, where non-trivial particle production takes place.
Moreover, the presence of non-zero vacuum energies, and in general of non-vanishing tensor components of the stress tensor of the fermion fluid, indicates breaking of Lorentz invariance, except in the case of
de-Sitter (or anti-de-Sitter ) vacua, with an equation of state between pressure $p$ and energy density $\rho$ of the form $p=-\rho$. For the flavour-vacuum, as shown in \cite{conundrum}, and mentioned above,  the fermion field theory with mixing  leads to equations of states that depend on the Universe era, with $w \to -1$ approximately only at late eras. We should note that, in our opinion, this latter statement has only been argued  but not rigorously proved in \cite{conundrum}, since the relevant calculations have been performed in Minkowski flat space-time.
\begin{figure}[ht]
\centering
\includegraphics[width=7.5cm]{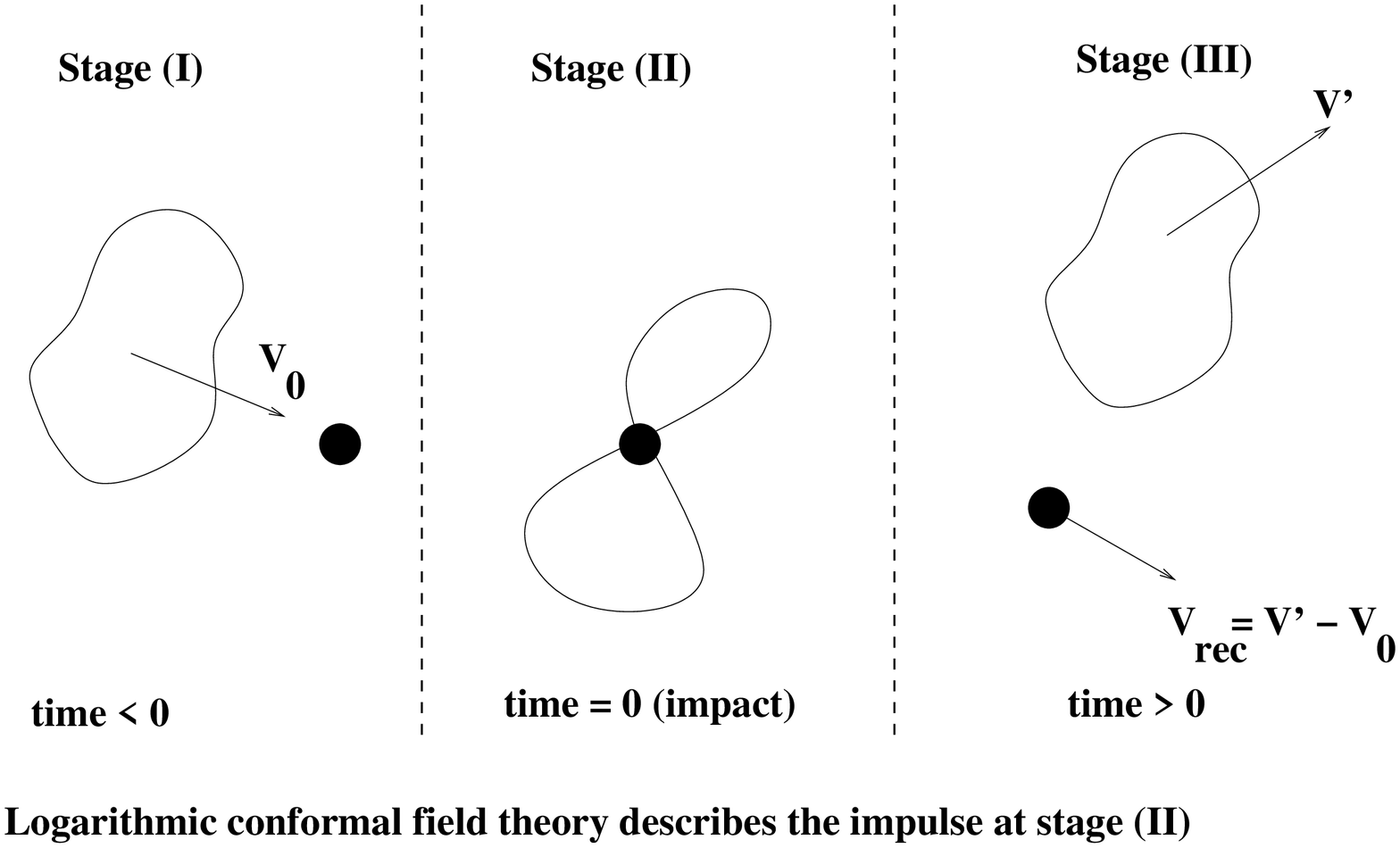} \hfill
\includegraphics[width=7.5cm]{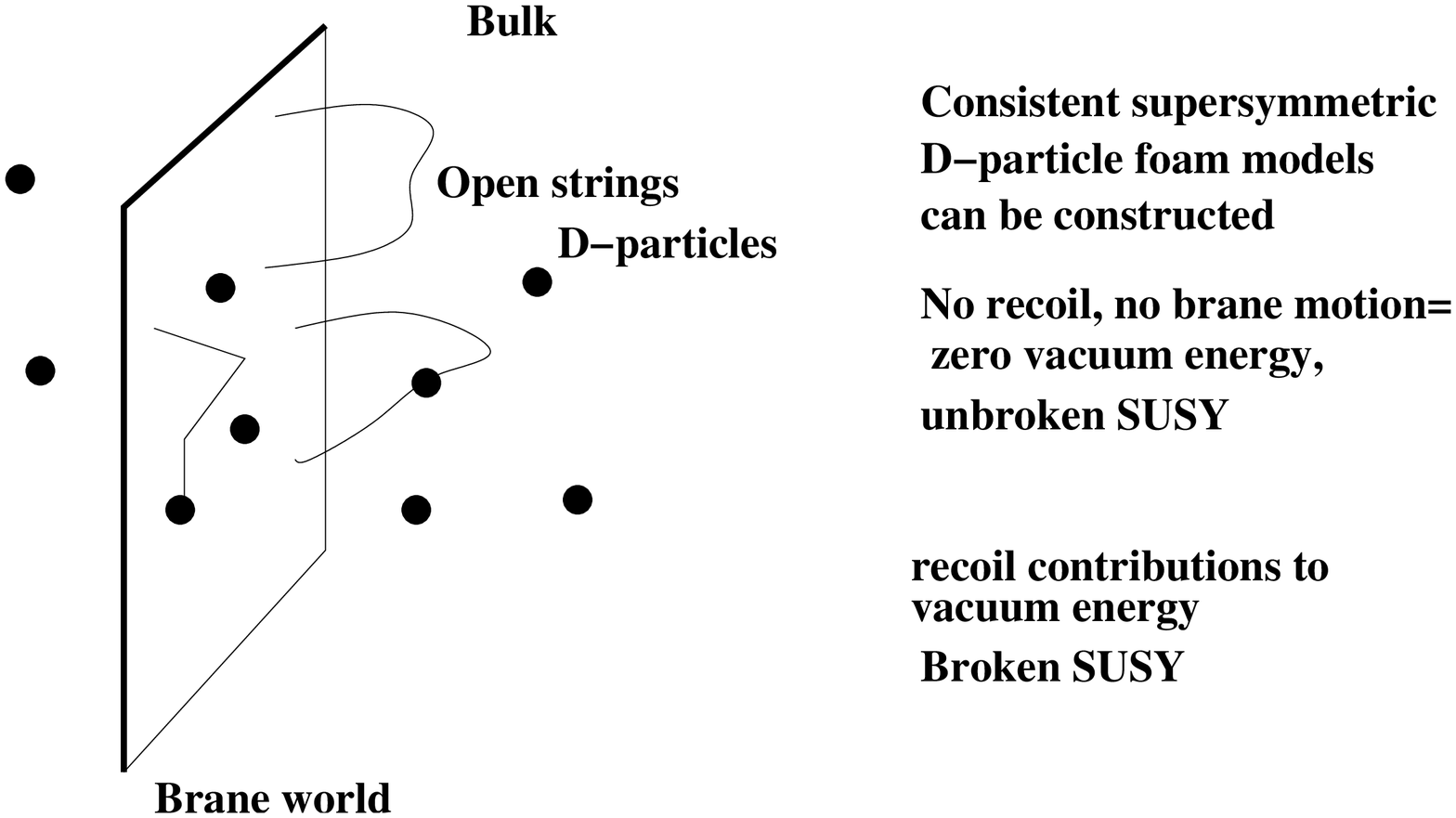} \caption{Schematic
representation of a D-particle space-time foam model. The figure indicates also the capture/recoil
process of a string state by a D-particle defect for closed (left picture) and open
(right picture) string states, in the presence of D-brane world. The presence of a
D-brane is essential due to gauge flux conservation, since an isolated
D-particle cannot exist. The intermediate composite state at $t=0$, which has
a life time within the stringy uncertainty time interval $\delta t$, of the
order of the string length, and is described by world-sheet logarithmic
conformal field theory, is responsible for the distortion of the surrounding
space time during the scattering, and subsequently leads to induced metrics
depending on both coordinates and momenta of the string state. This results on
modified dispersion relations for the open string propagation in such a
situation~\cite{Dfoam}, leading to \emph{non-trivial
optical properties} (refractive index\emph{ etc.}) for this space time.}%
\label{fig:recoil}%
\end{figure}
A first such step towards the construction of microscopic models that would provide mathematically and physically consistent realizations of the flavour vacuum, has been performed in \cite{mavrosarkar},
 in the context of the so-called D-particle foam model~\cite{Dfoam}, a string/brane inspired model of
 space-time foam. According to this model, our Universe, after perhaps appropriate compactification, is represented  as a three brane, propagating in a bulk space-time punctured by D0-brane (D-particle) defects. As the D3-brane world moves in the bulk, the D-particles cross it, and for an observer on the D3-brane the situation looks like a ``space-time foam'' with the defects ``flashing'' on and off (``D-particle foam''). The open strings, with their ends attached on the brane, which represent matter in this scenario, can interact with the D-particles on the D3-brane universe in a topologically non-trivial manner, involving splitting and capture of the strings by the D0-brane defects, and subsequent re-emission of the open string state (see fig.~\ref{fig:recoil}). However, the flavour of the re-emitted state may not be the same as that of the incident one, thereby leading to vacuum-induced flavour oscillations and mixing. It should be emphasized that, due to electric charge conservation, only \emph{electrically neutral} matter interacts non-trivially with the D-particle ``foam'', which is transparent to charged matter~\cite{ems,emnnewuncert}.

In such a model, the flavour vacuum is regarded as an effective description of the vacuum state for low-energy string modes with mixing, as a result of the  breaking of Lorentz symmetry locally in space-time, due to D-particle recoil during the string-D-particle interactions. Nevertheless, Lorentz symmetry is preserved on the average, in the sense that the appropriate vacuum-expectation-values (v.e.v.) of the relevant Lorentz-breaking observables vanish. However, this is not the case for the quantum fluctuations of those observables, which may be non-trivial.
We consider the first-quantised string-theory framework, that describes (perturbatively) the physics of (matter) open strings either stretched between the D-particle and the D3-brane world, with both their ends attached to the D3-brane (c.f. fig.~\ref{fig:recoil}); quantum fluctuations of target-space background fields, in which the string propagates, are induced by appropriate summation over world-sheet surfaces with higher topologies (genera). In this sense, a quantum fluctuating D-particle in the foam, will be described by such stretched open strings, with at least one of their ends attached to it.

The structure of the article is as follows: in the next section \ref{sec2} we review briefly the D-particle foam model and discuss the formalism that leads to an induced space-time metric of the form of a conformally flat expanding Universe, as a result of the space-time fluctuating D-particle background.
In section \ref{sec3} we discuss a \emph{gravitational MSW} effect as a result of the existence of the D-foam, that leads to gravitational-medium-induced flavour mixing and mass differences. We also present in that section plausibility arguments for the r\^ole of the flavour Fock vacuum and the corresponding excitation states as the physical states in the problem. In section \ref{sec4}, in the context of a low-energy field theory limit, we discuss the contributions of bosonic degrees of freedom
to the dark energy of the brane universe, defined as the appropriate vacuum expectation value of the stress-energy tensor between flavour-Fock vacuum states.
In section \ref{sec5} we repeat the construction for the fermionic low-energy degrees of freedom, and evaluate the
relevant contributions to the dark energy of the brane. An important issue arises regarding the choice of the appropriate normal ordering that should lead to the physically correct subtraction of the (field-theoretic) ultraviolet divergences, in a way consistent with the gravitational MSW effect. This is discussed in detail in section \ref{sec6}. The reader's attention is called at this point to the fact that in string theory there are no actual ultraviolet momentum divergencies. These are artifacts of the low-energy local effective field theory which is defined up to energies of the order of the string scale, $M_s$, or better the Planck scale (defined as the ratio of $M_s/g_s$, with $g_s$ the string coupling, assumed weak $g_s < 1$). Thus, by definition any momentum integral will be cut-off
at that scale automatically. The subtraction procedure we are applying to the bosonic or fermionic stress-energy tensors in this field-theoretic context, defines the appropriate effective field theory degrees of freedom, accessible to a low-energy observer, and is consistent with the fact that any contributions to the vacuum energy should vanish in the absence of D-particle foam effects. In section \ref{sec7} we discuss the emergence of a momentum cut-off much lower than the Planck scale that arises from statistical arguments related to particle production that characterises our expanding background. This is not a sharp cutoff, but rather defines the appropriate physical degrees of freedom that lead to significant contributions to the brane-world vacuum energy. We also discuss in this section the equation of state of the fermionic vacuum condensate and demonstrate that, for late eras of the Universe, it behaves as a fluid with negative pressure and positive energy; however the equation of state has $w_{\rm fermion} > -1/3$ and so the contribution of the fermion-fluid flavour vacuum  alone could not yield accelerating Universes.  However, 
on taking into account the contributions to the vacuum energy coming from flavoured bosons, which are natural from the point of view of the
(broken) target-space supersymmetry that characterises realistic superstring/supermembrane models of space-time foam,
and which should be considered as (algebraically) additive to the fermion contributions, one may obtain the conditions for late-era acceleration of the Universe. Indeed, the boson fluid is also characterised by positive energy and negative pressure, but its equation of state is, for late eras, close to $w_{\rm boson} \to -1$ and hence overall the D-foam universe appears accelerating at late eras.
Finally, section \ref{sec8} contains our conclusions and outlook.

\section{D-particle foam induced metric \label{sec2}}

The target-space quantization of the recoil velocity of a D-particle, $u_i$, during its interaction with a matter open string is achieved~\cite{szabo} by a genus summation on the world-sheet, which in the case of a bosonic $\sigma$-model with a D-particle recoil deformation can be cast in a closed form~\cite{szabo,tsallisarkar}. This yields a stochastic Gaussian distribution of the recoil velocities $u_i$, around a zero average, with a variance $\sigma^2$ that depends at most on target time, and not on the position of the D-particle:
\begin{equation}
< u_i > = 0 ~, \quad < u_i u_j > = \sigma^2 \delta_{ij}~, \quad \sigma^2 \sim g_s^2 t_0^2~,
\label{gauss}
\end{equation}
where the time $t_0$ extends over the capture time of the string by the fluctuating D-particle.
As discussed in \cite{tsallisarkar,emnnewuncert}, for a matter string with total energy $p_0$, the capture time is of order $t_c \sim \alpha ' p^0$. Moreover in \cite{tsallisarkar} we also argued, and shall review below,
that there is a dynamically imposed upper bound scale (cutoff) for the momenta of the particle excitations interacting with the D-particles, which is of order of the mass of the particles, essentially (c.f. (\ref{bosoncutoff}), below). Even if one considers sneutrinos, their masses (as a result of supersymmetry breaking) may be assumed of a few TeV in phenomenologically interesting models, which is still much smaller than a Planck-scale mass of a D-particle. Hence the capture time $t_0$ is small for our flavoured cases examined here. One has to average the relevant expressions yielding pressure and energy density of the particle fluid in the flavour-vacuum over such time scales. Because the time scales involved are small, such time averages may be replaced, at a good approximation, by the value of the relevant quantity to be averaged over a time scale $t_0$.
This means that on a global scale, on the D3 brane Universe, the quantum fluctuating D-particles will fluctuate with an average variance $\sigma^2(t_0) \ll 1$. On using a dilute gas approximation we can take a statistical average over populations of D-particles  whose \textit{density} does depend on the cosmological era of the Universe; the D-particle foam could then produce an isotropic and homogeneous (cosmological type) space-time background,
with non-trivial particle production, in which flavour mixing takes place in a self-consistent way~\cite{mavrosarkar}, as we shall review below.

For the benefit of the reader, we feel it would be instructive to first review briefly the mathematical formalism underlying the quantum-fluctuating D-particle ``foamy'' space time.
The world-sheet boundary operator $\mathcal{V}_{\rm{D}}$ describing the
excitations of a moving heavy D0-brane is given in the tree approximation by:
\begin{equation}
\mathcal{V}_{\rm{D}}=\int_{\partial D}\left(  y_{i}\partial_{n}X^{i}%
+u_{i}X^{0}\partial_{n}X^{i}\right)  \equiv\int_{\partial D}Y_{i}\left(
X^{0}\right)  \partial_{n}X^{i} \label{recoilop}%
\end{equation}
where $\partial D$ denotes the boundary of the world-sheet $D$ with the topology of a disk, to lowest order in string-loop perturbation theory,
$u_{i}$ and $y_{i}$ are the velocity \ and position of the heavy (non-relativistic) D-particle
respectively and $Y_{i}\left(  X^{0}\right)  \equiv y_{i}+u_{i}X^{0}$. To
describe the capture/recoil we need an operator which has non-zero matrix elements between
different states of \ the D-particle and is turned on ``abruptly'' in target time. One way of doing this is to put~\cite{kogan} a $\Theta\left(  X^{0}\right)  $, the Heavyside function, in front of
$\mathcal{V}_{\rm{D}}$ which models an impulse whereby the D-particle starts
moving at $X^{0}=0$. This impulsive $\mathcal{V}%
_{\rm{D}}$, denoted by $\mathcal{V}_{\rm{D}}^{imp}$, can thus be represented
as
\begin{equation}
\mathcal{V}_{\rm{D}}^{imp}=\frac{1}{2\pi\alpha '}
\sum_{i=1}^{d}\int_{\partial D}d\tau\,u_{i}%
X^{0}\Theta\left(  X^{0}\right)  \partial_{n}X^{i}. \label{fullrec}%
\end{equation}
where $d$ in the sum denotes the appropriate  number of spatial target-space dimensions.
For a recoiling D-particle confined on a D3 brane, $d=3$.

Since $X^{0}$ is an operator it will be necessary to define $\Theta\left(
X^{0}\right)  $ as a \emph{regularized}  operator using the contour integral%
\begin{equation}
\Theta_{\varepsilon}\left(  X^{0}\right)  =-\frac{i}{2\pi}\int_{-\infty
}^{\infty}\frac{d\omega}{\omega-i\varepsilon} \, e^{i\omega X^{0}} \quad \rm{ with } \, \quad\varepsilon
\rightarrow 0^+~,
\end{equation}
where $\varepsilon$ is a regulator, which, as discussed in \cite{kogan} and will be reviewed below, is linked with a running cutoff scale on the world-sheet of the string, on account of the requirement of the closure of the (logarithmic) conformal algebra.
Hence we can consider%
\begin{equation}\label{Depsilonop}
D_{\varepsilon}(X^0) = X^{0}\Theta_{\varepsilon}\left(  X^{0}\right)
=-\int_{-\infty}^{\infty}\frac{d\omega}{\left(  \omega -i\varepsilon\right)  ^{2}%
}\, e^{i\omega X^{0}}~.
\end{equation}
The presence of a recoil deformation leads to local distortions in the neighboring space time geometry which can be found as follows:
let one write the boundary recoil/capture operator $\mathcal{V}_{\rm{D}}^{imp}$ (\ref{fullrec}) in the Dirichlet picture as a total derivative over the bulk of the world-sheet by means of the two-dimensional version of Stokes theorem (omitting from now on the explicit summation over repeated $i$-index, which is understood to be over the spatial indices of the D3-brane world of fig.~\ref{fig:recoil}):
{\small \begin{eqnarray}\label{stokes}
&& \mathcal{V}_{\rm{D}}^{imp}=\frac{1}{2\pi\alpha '}
\int_{D}d^{2}z\,\epsilon_{\alpha\beta} \partial^\beta
\left(  \left[  u_{i}X^{0}\right]  \Theta\left(  X^{0}\right)  \partial^{\alpha}X^{i}\right) = \nonumber \\
&& \frac{1}{4\pi\alpha '}\int_{D}d^{2}z\, (2u_{i})\,\epsilon_{\alpha\beta}
 \partial^{\beta
}X^{0} \Bigg[\Theta_\varepsilon \left(X^{0}\right) + X^0 \delta_\varepsilon \left(  X^{0}\right) \Bigg] \partial
^{\alpha}X^{i}\nonumber \\
\end{eqnarray}}where $\delta_\varepsilon (X^0)$ is an $\varepsilon$-regularised $\delta$-function.
This is equivalent to a deformation describing an open string propagating in an antisymmetric  $B_{\mu\nu}$-background corresponding to an external constant in target-space ``electric'' field,
\begin{equation}
B_{0i}\sim u_i ~, \quad B_{ij}=0~,
\label{constelectric}
\end{equation}
where the $X^0\delta (X^0)$ terms in the argument of the electric field yield vanishing contributions in the large time limit $\varepsilon \to 0$, and hence are ignored from now on. We remark for completeness at this stage that, upon a T-duality canonical transformation of the coordinates~\cite{tdual}, the presence of the B-field leads to mixed-type boundary conditions for open strings on the boundary $\partial \mathcal{D}$  of world-sheet surfaces with the topology of a disc:
\begin{equation}
      g_{\mu\nu}\partial_n X^\nu + B_{\mu\nu}\partial_\tau X^\nu |_{\partial \mathcal{D}} = 0~,
\label{bc}
\end{equation}
with $B$ given by (\ref{constelectric}). Absence of a recoil-velocity $u_i$-field leads to the usual Neumann boundary conditions, while the limit where $g_{\mu\nu} \to 0$, with $u_i \ne 0$, leads to Dirichlet boundary conditions.

As discussed in detail in refs.~\cite{sussk1,seibergwitten},
 there is also an induced open-string \emph{effective target-space-time metric}.
To find it,
one should consider the world-sheet propagator on the disc $\langle X^\mu(z,{\overline z})X^\nu(0,0)\rangle$, with the boundary conditions (\ref{bc}).
Upon using a conformal mapping of the disc onto the upper half plane
with the real axis (parametrised by $\tau \in R$) as its boundary~\cite{seibergwitten},
one then obtains:
\begin{equation}
     \langle X^\mu(\tau)X^\nu(0)\rangle = -\alpha ' g^{\mu\nu}_{\rm open,~electric}{\rm ln}\tau^2 + i\frac{\theta^{\mu\nu}}{2}\epsilon(\tau)
\label{propdisc}
\end{equation}
with the non-commutative parameters  $\theta^{\mu\nu}$ given by
\begin{equation}
[ X^1, t ] = i \theta^{10} ~, \qquad \theta^{01} (= - \theta^{10}) \equiv \theta =  \frac{1}{u_{\rm c}}\frac{\tilde u}{1 - \tilde{u}^2}
\label{stnc}
\end{equation}
where $t$ is the target time; for definiteness the recoil can be assumed to be along the spatial $X^1$ direction, \emph{i.e.} $0 \ne k_1 \equiv k \parallel u_1~, k_2=k_3 =0$. $\epsilon(\tau)$ is the step function having value $-1$ for ${\tau}<0$ and $1$ for ${\tau}>0$. The quantity $\tilde{u}_i \equiv \frac{u_i}{u_{\rm c}}$ and  $u_{\rm c} = \frac{1}{2\pi \alpha '}$ is the Born-Infeld \emph{critical} field. Since the space and time co-ordinates are world-sheet fields, this commutator is calculated~\cite{seibergwitten} using the appropriate first-quantised string commutation relations on the world sheet.
The effective Finsler-type open-string metric~\cite{finsler}, due to the presence of the recoil-velocity field $\vec{u}$ (whose direction breaks target-space Lorentz invariance) is given by:
\begin{eqnarray}
           g_{\mu\nu}^{\rm open,electric} &=& \left(1 - {\tilde u}_i^2\right)\eta_{\mu\nu}~, \qquad \mu,\nu = 0,1 \nonumber \\
           g_{\mu\nu}^{\rm open,electric} &=& \eta_{\mu\nu}~, \mu,\nu ={\rm all~other~values}~.
\label{opsmetric}
\end{eqnarray}
There is, moreover, a modified effective string coupling~\cite{seibergwitten,sussk1}:
\begin{equation}
   g_s^{\rm eff} = g_s \left(1 - \tilde{u}^2\right)^{1/2}~.
\label{effstringcoupl}
\end{equation}
It should be mentioned here that such metrics have been suggested in the context of a T-dual Neumann picture~\cite{szabo} of the D-particle recoil process in refs.~\cite{Dfoam}.

The presence of a critical ``electric'' field is associated with a singularity of both the effective metric as well as the non-commutativity parameter and there is also an effective string coupling, which  vanishes in that limit (\ref{effstringcoupl}). This reflects the \emph{destabilization of the vacuum} when the ``electric'' field intensity approaches the \emph{critical value}, which was noted in \cite{burgess}. Since in our D-particle foam case, the r\^ole of the `electric' field is played by the recoil velocity of the D-particle defect, the critical field corresponds to the relativistic speed of light. This accords with special relativistic kinematics, which is respected in string theory, by construction.
The critical recoil velocity is of order one, which in turns sets the highest order of magnitude of energies of the stretched strings to the mass of the D-particle, $M_s/g_s$, as announced previously.
In this sense, the variances (\ref{gauss}) are characteristic constants, depending on microscopic parameters, such as the mass of the D-particles.

In the absence of any special knowledge, the most natural choice is to consider \emph{isotropic} cases of \emph{foam}, in which the induced Finsler metric assumes the conformal form
\begin{eqnarray}
           g_{\mu\nu}^{\rm open,electric} &=& \left(1 - {\tilde u}_i^2\right)\eta_{\mu\nu}~,
\label{opsmetric2}
\end{eqnarray}
for all $\mu, \nu =0, \dots 3$. This is the case we shall henceforth concentrate on.

Stochastic quantum fluctuations of the recoil velocity $u_i$, induced by the
summation over genera on the world-sheet~\cite{szabo}, imply, on account of (\ref{gauss}), a constant
metric of the form:
\begin{equation}\label{metric3}
  g_{\mu\nu} = \left(1 - \sigma^2\right)\eta_{\mu\nu}
 \end{equation}
 In addition to the quantum fluctuations of the metric as a result of a single D-particle
 fluctuation considered hitherto, one should consider the effects of a statistical population of D-particles, which characterises realistic cases of D-particle foam. Denoting such statistical averages over populations of D-particles (as opposed to quantum averages over fluctuations of a single quantum D-particle) by $\ll \dots \gg$, one should bear in mind that in general, in situations like in fig.~\ref{fig:recoil}, the density of D-particles in the bulk, which essentially the statistical average depends upon, may vary with the cosmological time scale, in the sense that their bulk distribution may not be uniform.
 Hence, on taking the statistical average of the metric (\ref{metric3}) over populations of D-particles,
 we obtain, in general, a time dependent induced metric:
 \begin{eqnarray}\label{confmetrtime}
  \ll g_{\mu\nu} \gg \equiv g_{\mu\nu}^{\rm stat} = \left(1 - \ll \sigma^2\gg(t)\right)\eta_{\mu\nu}~,
 \end{eqnarray}
 where $\ll \sigma^2\gg$(t) depends in general on the cosmological time, $t$, for reasons stated above.
 In fact, from the conformal nature of the induced cosmological metric (\ref{confmetrtime}), we see that the time $t$ that appears naturally in our construction is the so-called conformal time $\eta$ in standard Cosmology, and indeed the metric $g_{\mu\nu}^{\rm stat}$ acquires the standard cosmological form in the conformal time frame:
 \begin{equation}\label{cosmometr}
 g_{\mu\nu}^{\rm stat} = \C(\eta) \eta_{\mu\nu}~,
 \end{equation}
  where the scale factor of the D-foam universe is $C(\eta ) = 1 - \ll \sigma^2 \gg (\eta)$.
 Slightly expanding universes are obtained for cases in which the D3 brane moves towards a region in the bulk space (c.f.~fig.~\ref{fig:recoil}) characterised by a \emph{depletion} of D-particles. Notice that in our construction of small recoil velocities, $|u_i | \ll 1$, for which our perturbative $\sigma$-model treatment of strings suffices, the induced metrics are only slightly deviating from Minkowski space time, and as such they constitute good candidates to describe late-eras in our Universe'e expansion.
 This is the case we shall restrict ourselves on in this paper.

\section{D-particle Foam and Gravitational MSW Effect \label{sec3}}

Apart from the induced background space-time (\ref{cosmometr}),(\ref{confmetrtime}), the presence of a D-particle foam has other interesting consequences for matter flavoured states propagating on the D3 brane world.
Specifically, as advocated in \cite{barenboim}, the presence of a fluctuating ``medium'' of defects in the background space-time, may lead to induced mixing of flavoured states, and as a consequence to gravitationally-induced mass differences, analogous to the celebrated MSW effect~\cite{msw}. When neutrinos pass through ordinary matter media such as the Sun, the  mass differences and mixing angles acquire parts proportional to the (electronic) density of the medium according to the MSW effect. The difference in the (quantum) gravitational case is that the induced mass differences and mixing will now be proportional to the product of the density of fluctuating defects in space-time and Newton's constant $G_{N}$ (which expresses the coupling constant of gravity at an effective theory level). In terms of $M_P$, the four-dimensional Planck mass, we have $G_N = \frac{1}{M_P^2}$  (in units $\hbar = c = 1$) .
To be precise, it is argued in \cite{barenboim}, that the gravitationally-induced mass differences among flavour states will be of order
\begin{equation}
\Delta m^2_{\rm foam} \sim G_N \langle n_{\rm defect} \rangle p
\label{gravmsw}
\end{equation}
where $\langle n_{\rm defect} \rangle$ is an ``effective'' number density of the space-time defects (D-particles), probed by matter of momentum $p \equiv |\vec p|$. A measure of the weakness of the space-time foam is given by the smallness of the ratio $\frac{\Delta m^2_{\rm foam}}{{\overline m}^2}$ where ${\overline m}$ is a typical mass scale of the flavoured states. Indeed, for situations in which one has an effective number ${\cal N}^{\star}$ defects per Planck volume,
\begin{equation}
\Delta m^2_{\rm foam} \sim {\cal N}^{\star} \left(\frac{p}{M_P}\right)\,M_P^2  ~.
\label{gravmsw2}
\end{equation}
The value of ${\cal N}^{\star}$ takes into account scattering cross-sections of matter with the D-particles and so is much smaller than the number of D-particles per Planck volume. In the context of the string models we are considering, the four-dimensional Planck mass $M_P$ may be different from the string mass scale $M_s = 1/\sqrt{\alpha '}$, which is essentially a free parameter to be constrained by phenomenology, according to the modern approach to string theory.

To ensure $\Delta m^2_{\rm foam}$ has realistically small mass differences among neutrino flavours, i.e. do not exceed the observed values of order at most $10^{-3}$ eV$^2$,
one should have sufficiently dilute D-particle foams, such that ${\cal N}^{\star} \ll 1$.
This will be assumed throughout this work. A plausible assumption made in \cite{barenboim}, which, we
shall also adopt here, is that the induced mass differences are essentially independent of the momentum of the probe, since the effective density of defects decreases with the momentum $p$ (the faster the probe, the less time it has to interact with the foam in the MSW framework). One should notice, however, that such
an assumption is strictly necessary only if the foam-induced
mass differences are to account for the entire observed mass differences, which are indeed independent of the momentum of the neutrinos. This is unnecessary in the case here, since the foam-induced mass differences are only a tiny part of the experimentally measured ones, as required by the currently accepted experimental facts~\cite{barenboim}. Nevertheless, due to the smallness of $\Delta m_{\rm foam}^2$, compared to the typical neutrino mass scales, $\overline{m}$, it is convenient to ignore such momentum dependence since it suffices to effect the order-of-magnitude estimates in this article. This will be assumed from now on.

It is important to note that the gravitational MSW effect pertains to flavour mixing induced by the medium and does not induce any distortion of space-time \emph{per se}. According to our discussion in the previous section, it is the recoil of the fluctuating defect that induces the background (\ref{confmetrtime}).
This is an important distinction that should be used later on, when we discuss ultraviolet subtractions in our effective low-energy theory of string matter interacting with the defects. However, both effects are affected by the density of defects.

In our model the flavour mixing originates from the fact that, during the capture process of an open string (matter) state by the
D-particle, the mass $m$ of the re-emitted state might be different from the incident one.
In this sense, the D-particle ``medium'' will induce flavour oscillations and mixing, and as a consequence the flavour Fock space vacuum is appropriate for quantization of those states, since the ``flavour'' states are the appropriate physical states in this context~\cite{mavrosarkar}.
Unfortunately at present, our understanding of such a mechanism from a superstring-model, such as type IIB string theory, discussed by Li \emph{et al.} in \cite{emnnewuncert},  is inadequate, due to the non-perturbative nature of the process of D-particle-induced mass flips, in a region of strong gravity.
Such processes require knowledge of the dynamics of D-particles \emph{per se}, and, unlike the simple recoil/capture processes that do not involve mass changes,
cannot be simply described by means of (perturbative) world-sheet methods.
Nevertheless the time scale involved in such a mass flip can be estimated, using stringy uncertainty relations that are  independent of the details of the underlying microscopic string model.
We first note that when a D-particle interacts with a pair of open string states stretched between the defect and the D3 brane, representing the capture and splitting  process (as in fig.~\ref{fig:recoil}), there is an induced repulsive short-range potential ${\cal V}$, calculated by means of appropriate world-sheet annulus graphs in \cite{Dfoam2},  following techniques developed in \cite{annuli}. The relevant parts of the potential for our discussion in this paper are of the form
\begin{equation}\label{dod3pot}
{\cal V} \ni -\frac{\pi \, \alpha '}{12}\frac{u^2}{r^3}~,
\end{equation}
where $u$ is the relative four-velocity (i.e. $u^2 = \frac{v^2}{1-v^2}$ in units of the speed of light \emph{in} (Minkowski) \emph{vacuo} $c=1$ and $v$ is the 3-velocity) between the D-particle and the D3-brane and $r$ is the distance between the defect and the brane along the transverse directions to the brane world (c.f. fig.~\ref{fig:recoil}; there is no potential for D-particle motion parallel to the brane~\cite{annuli})~\footnote{In the supersymmetric model of \cite{Dfoam2}, the brane is a D8-brane, from which a D3-brane can be obtained by appropriate compactification. The form of the (repulsive) short-range interaction, and in particular the dependence on $u$ and $r$, are insensitive to such  compactification details, but do depend on the dimensionality of the interacting branes~\cite{annuli}. On the other hand, the precise form of the numerical coefficient depends on the details of the construction and on the presence of other branes and orientifold planes (the latter ensuring dynamical compactification of the bulk space in the model). However, for our order of magnitude estimates such issues are not relevant.
Moreover, the presence of orientifold planes in the construction of \cite{Dfoam2} can cancel velocity independent terms in the potentials.
Such terms are also irrelevant since we are only interested in potential fluctuations induced  by the velocity fluctuations, see eq.~(\ref{vflct}) below. }.  In our case, with a fundamental string stretched between the two, the order of this distance can vary typically from
that of the string length $\sqrt{\alpha '}=1/M_s$ to a (much smaller) characteristic minimum one $L_{\rm min} \sim {\overline{m} } \alpha ' \ll \sqrt{\alpha '}$ where $\overline{m}$ is a typical mass scale of the stretched light string state representing the flavoured states~\cite{annuli, Dfoam2}. The corresponding magnitudes $|{\cal V}|$  are $|{\cal V}|  \sim M_s \,u^2$ and $|{\cal V}|  \sim \left(\frac{M_s}{\overline{m}}\right)^3 \, M_s \,u^2$.
In this work we will restrict our attention to the simple case of two dominant mass eigenstates, with masses $m_i, i=1,2$; one may then take $\overline{m} \sim \frac{1}{2}(m_1 + m_2)$.
It will be convenient to collectively represent the effects of these two extreme cases for the potential in a single formula:
\bea
|{\cal V}| \sim \left(\frac{M_s}{\overline{m}}\right)^q \, M_s \,u^2~,
\label{abspot}
\eea
with $q=0 (3)$ for the former (latter) case.

A D-particle induced mass flip of a flavoured matter string state, such as a neutrino,  will be between masses separated by terms of order (\ref{gravmsw}), (\ref{gravmsw2}) in our scenario. Since  the recoil contribution of the D-particle cancels out on average, one has correspondingly a momentum conservation for the neutrino state interacting with the defect.
In the \emph{non-relativistic limit} of the D-particle recoil velocities $|u| \simeq |v| \ll 1$ energy conservation and in a first quantized weakly coupled $g_s < 1$ string theory framework, energy conservation implies:
\be
M_D + \ll (p^2 + {m_1}^{2})^{1/2}  \gg = \ll (p^2 + {m_2}^{2} )^{1/2} \gg + M_D + \frac{1}{2}\,M_D \ll u^2 \gg + {\cal O}(u^4) ~,
\label{encons}
\ee
where $M_D = \frac{M_s}{g_s} \sim M_P$ is the D-particle mass (assumed to be of order of the four-dimensional
Planck mass $M_P$). This allows us to estimate the recoil velocity fluctuations during a mass-flip process. The above relation expresses an average over \emph{both} quantum fluctuations of the recoil velocity at the individual D-particle level and over D-particle foam populations. This averaging is denoted by $\ll \dots \gg$.
On denoting the neutrino energy difference by $\Delta E $, one may estimate from (\ref{encons}):
\begin{equation}\label{encons2}
M_s \ll u^2 \gg \sim 2 g_s \, \Delta E ~.
\end{equation}
Hence, the recoil velocity fluctuation accompanying a mass-flip induces in turn a fluctuation in the potential (\ref{dod3pot})
of order (for $r \sim \left(\overline{m}\sqrt{\alpha '}\right)^{q'}\sqrt{\alpha '}$, with ${q'}=0$ or $1$,  corresponding to the two characteristic scales of the intermediate string states discussed above) :
\begin{equation}
\Delta {\cal V} \sim 2 \left(\frac{M_s}{\overline{m}}\right)^q g_s \, \Delta E~.
\label{vflct}
\end{equation}

In the laboratory frame from the saturation of the energy-time uncertainty relation an energy fluctuation will imply a life-time $\Delta t^q_{\rm mass-flip}$  for the intermediate string state associated with the mass flip. It suffices to consider the low-energy quantum  mechanical version of this uncertainty, which yields (in units $\hbar = 1$)
\begin{equation}\label{tflip}
\Delta t^q_{\rm mass-flip} \sim \frac{1}{\Delta{\cal V}} \sim \left(\frac{\overline{m}}{M_s}\right)^q\,\frac{1}{2 g_s \, \Delta E}~, \qquad q=0, 3~.
\end{equation}
We will denote by $\Delta t_{\rm capture}$ the time during which an intermediate string state, stretched between the D-particle and the D3-brane world, grows from zero size to its maximal one permitted by the stringy time-space uncertainty relations, and back to zero size. To have the possibility of mass-flip in string theory, $\Delta t^q_{\rm mass-flip}$ must be longer than the time involved in \emph{capture}, $\Delta t_{\rm capture}$.
As discussed in \cite{emnnewuncert},
the capture time is of order
\begin{equation}
\Delta t_{\rm capture} \sim \frac{\alpha ' p^0}{1 -u^2}~,
\label{tcapture}
\end{equation}
where $u^2$ is the D-particle recoil velocity, and
$p^0 \sim \left( p^2 + {\overline m}^2 \right)^{\frac{1}{2}}$ is the energy of the incident string state.

The delays (\ref{tcapture}) are consistent with  the time-space uncertainty relation $\Delta t \Delta X \sim \alpha '$, characteristic of string theory~\cite{yoneya} and actually \emph{saturate} it. They
can be computed rigorously within superstring theory in D-particle backgrounds by evaluating the relevant scattering amplitudes and looking at backward scattering contributions~\cite{sussk1,emnnewuncert}.
In what follows we shall only be interested in contributions to leading order in small quantities, and so, when using (\ref{tcapture}), the recoil velocity $u$ will be ignored.
The capture time (\ref{tcapture}) does not involve mass flip, and it is associated simply with
capture and re-emission of an open string state by the D-particle.

We reiterate that, in order for the mass-flip process to be feasible within a string theory model,
it is necessary that the capture time (\ref{tcapture}), which saturates the stringy time-spaced uncertainties, is \emph{shorter} than, or \emph{at most equal} to, the mass-flip time (\ref{tflip}). Otherwise, mass flip does not take place.
Such a requirement, then, implies an \emph{upper} bound for the spatial momenta of the states that can possibly undergo mass flip, i.e. for the momenta associated with the Fock-type flavour vacuum~\cite{vitiello,mavrosarkar}

First let us consider the case $q=0$.
It is easy to see from (\ref{tflip}) and (\ref{tcapture}) that
for all momenta that define an effective-low energy theory, i.e. momenta
smaller than the Planck scale $p < M_s/g_s$, the condition
\begin{equation}\label{cond}
\Delta t^q_{\rm mass-flip} > \Delta t_{\rm capture}
\end{equation}
is comfortably satisfied, since to violate it requires energy scales
\begin{equation}
E  > \frac{M_s}{g_s} \, \frac{M_s}{2\Delta E} \gg M_P \equiv \frac{M_s}{g_s}~.
\label{violcutoff}
\end{equation}
For a ratio $\xi$ of typical neutrino mass scale $\overline{m}$ to momenta, with $\xi$ either small or large, compared to unity, this inequality can only be satisfied for mass differences $\delta m^2 \equiv |m_1^2 - m_2^2|  > M_s^2/g_s$. This requirement is physically \emph{absurd}, and thus yields no physically sensible constraint on the neutrino mass differences for the case $q=0$.

However, on taking into account the case $q=3$, the condition (\ref{cond}) is satisfied for all momenta smaller than the cutoff-scale $M_s/g_s$ \emph{provided}
\begin{equation}\label{cond2}
\delta m^2 \le  \overline{m}^2 \,\left(\frac{\overline{m}}{g_s\,M_s}\right)
\end{equation}
which is a strong constraint for the foam-induced mass differences.

The bound (\ref{cond2}) leads to very small foam-induced mass differences for large string mass scales, $M_s$, close to the Planck scale $M_P$, while one can get mass differences of the order of the observed ones for $M_s$ of the order of TeV, and $g_s \sim 10^{-16}$, such that $M_s/g_s = M_P \sim 10^{19}$ GeV. Compactification details, of course, in phenomenologically realistic string/brane models may affect such estimates seriously. The reader should notice that in the bound (\ref{cond2}) there is a \emph{theoretical uncertainty} at most of order $\mathcal{O}(10)$, due to the \emph{uncertainty} in the  \emph{numerical coefficients} in the potential (\ref{dod3pot}), which depend on the details of the microscopic model, as already mentioned.

Hence mass flip processes, at least from the point of view of stringy uncertainties, are consistent physical processes that can take place for neutrino energy differences of physical relevance. A microscopic understanding of these processes in detailed realistic superstring/supermembrane models is, of course, still pending, and hence we can only give here plausibility arguments on the existence of such processes.

It should also be noticed that,
for small masses over momenta, and on assuming, for concreteness, only a D-foam-induced mass difference (\ref{gravmsw2}) among flavours, one may estimate from (\ref{encons}) the average stochastic fluctuations of the D-particle foam recoil velocities:
\begin{equation}
\ll u^2 \gg  \, = \,  \ll \sigma^2 \gg \simeq g_s \frac{\Delta m^2_{\rm foam}}{M_s \, p} \simeq
{\cal N}^{\star} \ll 1 ~.
\label{u2msw}
\end{equation}
Here we should remember that ${\cal N}^\star $ denotes the \textit{effective} number of D-particle defects contained in a Planck volume.
The reader should then notice the form similarity of (\ref{u2msw}) with the gravitational MSW-like relation (\ref{gravmsw}) conjectured in ref.~\cite{barenboim}. The order of the estimate, of course, may change significantly if the induced mass differences among neutrino flavours due to the foam constitute only a small percentage of the physically observed one ( as is most likely the case) even if the D-particle foam is physically relevant~\cite{barenboim}.

\section{Low-Energy Bosonic Field-Theory Mixing and Flavour Vacua \label{sec4}}

In this article, we shall discuss mixing of field theory excitations, induced on both bosonic~\cite{mavrosarkar} and fermionic excitations of strings by D-particles, during their topologically non-trivial  interactions (spiltting/capture/re-emission) with strings. However, the mixing will be discussed in the presence of a slightly expanding universe (\ref{confmetrtime}), induced on global scales by time varying populations of quantum fluctuating D-particles, as discussed above.
The presence of both fermionic and bosonic  ``flavoured'' field theory excitations of strings, finds a natural application in the case of superstrings and super-D-branes, which is the ultimate physical theory we have to consider. Indeed, even if supersymmetry is eventually broken in target space, the partners (e.g. sneutrinos) do exists, and in the case of flavour, their mass differences might be of the same order as the original particles (corresponding to flavoured neutrinos in our example), despite the fact that the mass differences between particles and their supersymmetry partners might be at least a few TeV, due to the broken supersymmetry.
The only caveat in our mathematical construction is that it is based on bosonic string theory, where the resummation of leading modular divergencies in the case of recoil-velocity deformed $\sigma$-models, is possible~\cite{szabo}. Unfortunately in the supersymmetric case, which would necessitate world-sheet supersymmetry as well, such a resummation is not possible at present~\cite{szabo2}. Thus for the (realistic) case of (broken) supersymmetric D-foam we could only assume that the conclusions drawn from the bosonic case, regarding stochastic fluctuations properties of the foam, are sufficiently robust to be extendable here.

We commence our discussion with a review of the bosonic case.
The (1+1)-dimensional bosonic case has been discussed in detail in \cite{mavrosarkar} and will not be repeated here, apart from pertinent information needed for completeness of our discussion.
It has been shown  there that the vacuum condensate in the case of scalar fields, taken as a representative example, behaves as a fluid with $w\approx -1$ once the MSW effect is taken into account.
The mixing/expansion is considered in a ($1+1$) dimensional framework, since the
recoil of the D-particle has been taken parallel to the motion of the bosonic excitations, assumed  along one spatial direction, say $X^1$ (c.f. (\ref{opsmetric}).
In view of our isotropic foam situation, considered here, (\ref{opsmetric2}), (\ref{confmetrtime}) our findings should carry forward to the full ($3+1$)
dimensional case.

For a scalar field $\phi$, the stress-energy tensor is:
\be
T^{bos}_{\mu\nu}[\phi]=\frac{1}{2}\left(D_{\mu}\phi D_{\nu} \phi +D_{\nu}\phi D_{\mu} \phi \right)-g_{\mu\nu}\lag
\ee
The $1+1$ dimensional version (with $g_{\mu\nu}=\C(\e)\e_{\mu\nu}$ in the conformal frame (c.f. (\ref{opsmetric})) was considered in ~\cite{mavrosarkar},
where it was shown that the only non-trivial components
are the diagonal ones:
\bea\label{stbos}
T^{bos}_{00}[\phi]&=&(\partial_{\e}\phi)^2+(\partial_{x}\phi)^2+\C_{eff}(\e)m^2\phi^2\nn
T^{bos}_{ii}[\phi]&=&(\partial_{x}\phi)^2+(\partial_{\e}\phi)^2-\C_{eff}(\e)m^2\phi^2
\eea
The reader should have noticed that we used the symbol $\C_{eff}$ in the mass term for the scalar
field and not simply $\C m^2 \phi^2$.  The quantity $\C_{eff}$
contains both the effects of the expansion and the MSW effect and  plays the r\^ole of an effective scale factor; this is in a similar spirit to standard effective field theories of inflation~\cite{chung} with interactions of massive dark matter particles to the inflaton field. The dispersion relations of such massive particles, and the associated stress tensor components, include the effective scale factors, as above. We shall come back to this point with more details when we discuss the fermionic case, where an entirely analogous situation applies.

The expressions (\ref{stbos}) readily
generalize to $3+1$ dimensions as follows:
\bea
T^{bos}_{00}[\phi]&=&(\partial_{\e}\phi)^2+\sum_{j=1}^3 (\partial_{x_j}\phi)^2+\C_{eff}(\e)m^2\phi^2\\
T^{bos}_{ii}[\phi]&=&(\partial_{\e}\phi)^2+\sum_{j=1}^3 (\partial_{x_j}\phi)^2-\C_{eff}(\e)m^2\phi^2 +2(\partial_{x_i}\phi)^2\nonumber
\eea
where the sums over spatial three-dimensional indices are explicitly denoted for clarity. Since the form of this term does not change when we go from the $1+1$ dimensional analysis to the $3+1$ one, the conclusion that $w\approx -1$ is valid. As discussed in \cite{mavrosarkar}, the appropriate normal ordering (subtraction) in our case of D-particle foam has to remove any terms that do not vanish in the limit where the variance of the D-particle fluctuations vanishes, $\ll~\sigma^2~\gg~\to~0$. We further assume that at late eras of the universe the D-particle fluctuations are weak, so only leading order terms in an expansion in powers of $\ll \sigma^2 \gg $ are kept.

Hence, after the appropriate subtraction, discussed further in \cite{mavrosarkar},
one arrives at
\bea\label{bostenssub}
\lfv :T^{bos}_{00}[\hat{\phi}]: \rfv&=&\lfv :\C_{eff}(\e)m^2\hat{\phi}^2:\rfv\nn
\lfv :T^{bos}_{ii}[\hat{\phi}]: \rfv&=&-\lfv :\C_{eff}(\e)m^2\hat{\phi}^2:\rfv~.
\eea
Any extra contributions acquired due to the higher ($(3+1)$) dimensionality, as compared to the $(1+1)$-dimensional case of ref.~\cite{mavrosarkar}, are \textit{common} to the two components; therefore the equation of state $w\approx-1$
\textit{holds} also in the $3+1$ dimensional bosonic field case.

To recapitulate, for weak D-particle foam, the energy density of the bosonic fluid in the flavour-vacuum formalism is positive, while the pressure is negative, and the equation of state is consistent with a cosmological constant. There is however an important issue to be addressed here. The expressions in (\ref{bostenssub}), which involve integration over momenta, are formally ultraviolet divergent \cite{mavrosarkar}. Hence in the low-energy field theory limit, the momentum integrals need a momentum scale cut-off $k_{\rm max}$.
It should be emphasised that in string theory there are no actual ultraviolet momentum divergencies. As already mentioned in the Introduction, these are artifacts of the low-energy local effective field theory which is defined up to energies of the order of the the Planck scale $M_s/g_s$. Thus, by definition any momentum integral will be cut-off
at that scale automatically. The appropriate effective field theory degrees of freedom, accessible to a low-energy observer, are defined by the subtraction procedure that we are applying to the bosonic or fermionic stress-energy tensors in this field-theoretic context. Moreover any such contributions to the vacuum energy should vanish in the absence of D-particle foam effects.
It is in this sense that a cut-off $k_{\rm max} $ is used in the model.

In \cite{mavrosarkar}, such a cutoff scale, which is, however, much smaller than the Planck mass, was determined dynamically, by considering particle production. The result of the $(1 + 1)$-dimensional case of \cite{mavrosarkar} has yielded the cutoff scale
\begin{equation}\label{bosoncutoff}
k_{\rm max} \sim \sqrt{\frac{m_{1}^{2} + m_{2}^{2}}{2}}~,
\end{equation}
where $m_{(i)},~i=1,2$ are bosonic eigenstate masses.
This is a result of the fact that the particle production falls off with the momentum, in such a way that the vacuum is populated significantly by flavoured bosons for momentum scales below (\ref{bosoncutoff}), and thus it is in such regimes of four-momenta that the condensate becomes significant.
For small mass differences, compared to masses, which we assume throughout our works~\footnote{We remark that, even if the bosons refer to sneutrinos, which have heavy masses due to target-space supersymmetry breaking, the relative mass differences between mass eigenstates may be assumed sufficiently small, since the mass differences are independent on supersymmetry, especially if, according to our D-particle foam model, these mass differences originate from foamy interactions, and hence are quantum-gravitational in origin.}, we may use the parametrization~\cite{mavrosarkar}:
\bea\label{bosons}
&& m^{(\pm)} = {\overline m} \pm \frac{\delta m}{2}~, \quad m^+ \equiv m_{1}, \quad m^- \equiv m_{2}
\nonumber \\
&& k_{\rm max} \sim {\overline m} +\frac{1}{8} \frac{(\delta m)^2}{{\overline m}}~, \nonumber \\
&& \delta m =  m_{1} - m_{2} \, \ll \, {\overline m}~.
\eea
Upon inserting the cutoff in the momentum integrals and performing the appropriate subtractions (\ref{bostenssub}), based on the metric (\ref{confmetrtime}), we arrive at the estimate
for the boson-induced flavour-vacuum energy density
\begin{equation}\label{bosons2}
   \rho_{\rm bosons} \sim {\rm sin}^2\theta \ll \sigma^2 (t_0)\gg \left(\delta m^2\right) ^2
\end{equation}
in the case of predominant two-flavour mixing, which we restrict ourselves here for concreteness and brevity.
The estimate requires time averaging of (\ref{bostenssub}) over small capture times, $t_0$, which in our model has been estimated to be close to zero~\cite{mavrosarkar}. This fact makes oscillatory terms, that may appear in the components of the stress tensor, negligible. Notice in our case the extra suppression factor $\ll \sigma^2 \gg (t_0)$, which for the cases of weak gravitational foam is smaller than one, as compared to the flat-space time flavour vacuum case of \cite{dark energy,conundrum}. In view of (\ref{bosons2}) bounds on $\ll \sigma^2 \gg$ may then be imposed by
cosmological considerations, given the order of magnitude of the dark energy at present eras of the Universe, observed today. It is important to note that, in our approach, we assume that the relative motion of D3-brane worlds in the bulk at present eras  is such that the associated supersymmetry breaking due to brane motion and the pertinent contributions to vacuum energy are negligible compared to the flavour-vacuum ones (\ref{bosons2}). This is an assumption that holds also for the fermionic contributions.

 \section{Low-energy Field-Theory Mixing and Fermionic Flavour-vacua in (3+1)-dimensions \label{sec5}}

 As we shall discuss below, as far as the equation of state is concerned, a crucial difference appears when one considers fermionic low-energy field theory excitations in the context of flavour vacua. When the normal ordering procedure is applied, the resulting equation of state is of the form
 $0 > w > -1/3$, a range insensitive to the cut-off. Thus, fermions alone, cannot lead to a present-epoch acceleration of the Universe through this mechanism. However, in our D-particle supersymmetric foam, where, for reasons stated, one has \emph{both } bosons and fermions as a result of \textit{broken} target-space supersymmetry, the contributions to the equation of state of the flavour vacuum from flavoured bosons (e.g. sneutrinos) can lead to a current era acceleration, by affecting the equation of state appropriately.

Before presenting the details of our calculation, let us briefly summarize
the formalism for the mixing of two fermionic (spin 1/2) flavours in a Minkowski space-time background:
two flavoured fermions $\psi_e(x)$ and $\psi_\mu(x)$ can be constructed from two free Dirac fields $\psi_1(x)$ and $\psi_2(x)$
with definite masses $m_1\neq m_2$, by means of the relation
\be
\left\{\begin{array}{rcl}
\psi_e(x)&=&\psi_1(x) \ct+\psi_2(x)\st\\
\psi_\mu(x)&=&-\psi_1(x) \st+\psi_2(x)\ct
\end{array}
\right.
\ee
It has been shown \cite{vitiello} that in quantum field theory (QFT) it is possible, in a finite volume, to define rigorously a unitary operator that behaves as the generator
of the mixing transformation for fields:
\be
\left\{\begin{array}{rcl}
\hat{\psi}_e(x)&=&\hat{G}_\theta^{\dagger}(t)\hat{\psi}_1(x)\hat{G}_\theta(t)\\
\hat{\psi}_\mu(x)&=&\hat{G}_\theta^\dagger(t)\hat{\psi}_2(x)\hat{G}_\theta(t)
\end{array}
\right.
\ee
This can be written as
\be\label{G}
\hat{G}_\theta(t)=\exp\left[\theta\! \int\! d\vx \left( \hat{\psi}^\dagger_1(x)\hat{\psi}_2(x)-\hat{\psi}_2^\dagger \hat{\psi}_1(x)\right)\right]~,
\ee
thus allowing the definition of a \textit{flavour vacuum} as the state
\be\label{fv}
\rfv\equiv \hat{G}_\theta^\dagger(t)\rmv
\ee
with $\rmv$ the mass eigenstate vacuum used in the quantization of the field theory of
$\hat{\psi}_1(x)$ and $\hat{\psi}_2(x)$. We should recall that in the thermodynamic, infinite-volume limit, as in the bosonic case,  the vacuum states $\rmv$ and $\rfv$ are orthogonal. Indeed this is the case for all the Fock-space excited states constructed out of these vacua~\cite{vitiello,ji}.  However, as discussed in \cite{mavrosarkar} and also below, in our string-inspired effective theory, flavour boson or fermion states, constructed out of $\rfv$, will exist up to a dynamically determined momentum cutoff scale. Above that scale, the pertinent states will be constructed out of the mass-eigenstate vacuum $\rmv$.  This is one way that our microscopic approach differs from that of \cite{vitiello,ji}.

We wish now to study how the \textit{flavour vacuum} expectation value of the stress-energy tensor varies in a weakly curved space-time background. For our model of D-particle foam, this is crucial, since the mixing phenomenon and a non-trivial, Robertson-Walker type, space-time background (\ref{confmetrtime}) are induced as a result of the quantum fluctuations of the D-particles. Thus, self consistency requires an analysis of the flavour vacuum in the presence of a non-trivial curved space-time, where particle production takes place.
To this end, it is of interest to first consider a free theory of two fields of definite mass in a curved space-time, that will be regarded as a classical
background. Then, under the assumption~\cite{mavrosarkar} that the metric is asymptotically flat at \emph{early} times, one can introduce the mixing (and therefore the \textit{flavour vacuum}) at $t\rightarrow -\infty$ in the way presented in the literature~\cite{vitiello}.
In our model of D-particle foam (c.f. fig.~\ref{fig:recoil}), the assumption of asymptotically flat space times at early times, $t \to -\infty$,
can be justified in a scenario in which the D3-brane worlds at $t \to -\infty$ finds itself in a bulk region depleted of D-particles. As the cosmological time elapses, the D3-brane world may move into a densely populated bulk regions of D-particles, and subsequently exit them at asymptotically long times in the future, $t \to \infty$, in such a way that an interpolating cosmological space-time between $-\infty$ and $+\infty$ may be induced as a result of such D-particle configurations. For current eras of the universe, we may assume that the density of D-particles is low enough so that there are only slight deviations from the Minkowski vacuum.

Adopting the Heisenberg picture~\cite{vitiello,mavrosarkar}, one can study the vacuum condensate
induced by the mixing. Hence the \textit{flavour vacuum} expectation value (\textit{vev}) of the stress-energy tensor operator is evaluated for a time-independent \textsl{flavour vacuum} state and a stress-energy tensor that \textsl{evolves} with time.
Therefore, we are going to construct the stress-energy tensor operator for a theory with two free fermions in a curved background and evaluate
its \textit{vev}, considering the \textit{flavour vacuum} defined in (\ref{fv}) and (\ref{G}) as \textit{the} vacuum, for $t\rightarrow -\infty$.

As has been well explained in the literature \cite{weinberg},
the classical theory for fields with spinorial structure is generalized in curved space-time through the
\textit{vierbein} formalism. The expression for the stress-energy tensor for a free $1/2$-spin field reads
\be\label{stress}
T_{\mu\nu}(\psi)=-g_{\mu\nu} \lag+\frac{1}{2}\left( \bar{\psi}\tilde{\gamma}_{\left(\mu\right.}D_{\left.\nu\right)}\psi-D_{\left(\nu\right.} \bar{\psi} \tilde{\gamma}_{\left.\mu\right)}\psi \right)
\ee
where $\lag$ denotes the Lagrangian of our theory,
$\tilde{\gamma}_{\mu}$ are the generalized $\gamma$-matrices defined by $\{\tilde{\gamma}_{\mu},\tilde{\gamma}_{\nu}\}=2g_{\mu\nu}$, $D_{\mu}$ is the gravitational covariant derivative and $\bar{\psi}\equiv i\psi \gamma^0$,
and $\gamma^0$ is the temporal component of the ordinary $\gamma$-matrices in the tangent plane (defined by $\{\gamma^{a},\gamma^{b}\}=2\e^{ab}$, $a,b=0,\dots 3$ tangent plane indices in the vierbein formalism). The stress-energy tensor for \textit{two} free fields
is simply $T_{\mu\nu}(\psi_1,\psi_2)=T_{\mu\nu}(\psi_1)+T_{\mu\nu}(\psi_2)$, therefore we can focus just on $T_{\mu\nu}(\psi)$.
Assuming that the flavour mixing does not affect the homogeneity and the isotropy of the universe we can consider a Friedmann-Robertson-Walker (FRW) metric in conformal coordinates $\{\e,\vx \}$,  $g_{\mu \nu}=\C(\e)\e_{\mu \nu}$, with $\C(\e)>0$.
In this case, adopting the convention $\e_{\mu\nu}=diag\{-1,1,1,1\}$, the stress-energy tensor becomes
\bea
 T_{\mu\nu}(\psi)=-\C\e_{\mu\nu}\lag +  \bpsi\!\left(\!\frac{\sqrt{\C}}{2} \gamma_{\left(\mu\right.}\partial_{\left.\nu\right)}
					  +\frac{\Cp}{16 \sqrt{\C}} \gamma_{\left(\mu\right.}[\gamma_0,\gamma_{\left.\nu\right)}]\!\right)\!\psi+ {\rm  h.c.}
\eea
An important remark on notation should be made at this point: from now on (i.e. in all subsequent formulae) the contraction of the space-time indices $\mu, \nu \dots$ is understood with respect to the Minkowski part of the metric, \emph{i.e.} ${\cal A}_\mu {\cal B}^\mu \equiv  \eta_{\mu\nu} {\cal A}^\mu {\cal B}^\nu $, since the scale factor $\C(\e)$ has been factored out appropriately. The notation regarding the contraction of tangent-space indices $a$, $b$ $\dots$, remains as before, the contraction involving the tangent-space Minkowski metric $\eta^{a b}$.

The quantization of this theory has been specifically carried out by \cite{parker} and discussed more generally by \cite{birrell}. According to \cite{parker}, the quantized spinor field can be written as
\bea\label{psiamin}
\hat{\psi}(\eta,\vec{x})=\left(\!\frac{1}{L \sqrt{\C(\eta)}}\!\right)^{\frac{3}{2}} \!\!\!\!\!\!
\sum_{\mbox{\tiny{$\begin{array}{c} \vp  \\ a,b=\pm 1 \end{array}$}}}  \!\!\!\!\!\!
\hat{a}^{(a,b)}(\vp,\e)
v^{(a,b)}(\vp,\e) \, e^{ia\left(\vp\cdot\vx-\!\int\!\!\w(\e) d\e\right)}
\eea
where $L$ is the parameter that enters the periodic boundary condition $\psi(\e,\vx+\vec{n}L)=\psi(\e,\vx)$ ($\vec{n}$ being a vector with integer Cartesian components),
 $\w(\e)\equiv\sqrt{p^2+m^2 \C(\e)}$,
$v^{(a,b)}(\vp,\e)$ is a spinor defined by
\be
\left\{
\begin{array}{l}
v^{(a,b)}(\vp,\e)\equiv v^{(a,b)}(\vp/\sqrt{\C(\e)})\\
(-i a\sqrt{p^2+m^2 } \gamma^0+i a \vec{\gamma}\cdot\vp+m)v^{(a,b)}(\vp)=0\\
v^{(a,b)\dagger}(\vp)v^{(a',b')}(\vp)=\delta_{a,a'}\delta_{b,b'}

\end{array}
\right.
\ee
and $\hat{a}^{(a,b)}(\vp,\e)$ are operators such that $\{\hat{a}^{(a,b)}(\vp,\e),\hat{a}^{(a',b')\dagger}(\vq,\e)\}=\delta_{a,a'}\delta_{b,b'}\delta_{\vp,\vq}$.
In order to introduce the mixing at early times, we first have to \textit{define} our Fock space for $\e\rightarrow -\infty$.
The relevant operators are $\hat{A}^{(a,b)}(\vp)\equiv\hat{a}^{(a,b)}(\vp,-\infty)$ with
$\{\hat{A}^{(a,b)}(\vp), \hat{A}^{(a',b')\dagger}(\vq)\}=\delta_{a,a'}\delta_{b,b'}\delta_{\vp,\vq}$.
Starting with these operators, we can define a Fock space for $\e\rightarrow -\infty$ following the usual prescriptions
of minkowskian QFT.
Moreover, if we assume that for our conformal scale factor behaves as $\C(-\infty)\rightarrow 1$,
we can introduce the mechanism of the mixing in the way that has been explained above.

Therefore the \textit{flavour vacuum} will be defined by
\be\label{fv2}
\rfv\equiv \hat{G}_{\theta}^{\dagger}(-\infty)\rmv
\ee
keeping in mind that, since the space-time is asymptotically flat at early times, the fields $\hat{\psi}$ behave in the usual relativistic way:
\be
\hat{\psi}(\eta,\vec{x})=\frac{1}{L^{\frac{3}{2}}} \!\!\!\!\!\!
\sum_{\mbox{\tiny{$\begin{array}{c} \vp  \\ a,b=\pm 1 \end{array}$}}} \!\!\!\!\!\!
\hat{A}^{(a,b)}(\vp)
v^{(a,b)}(\vp) e^{ia(\vp\cdot\vx-\sqrt{p^2+m^2}\e)}
\ee
for $\e\rightarrow -\infty.$  (N.B. the mass-eigenstate index $i$ is being omitted here for brevity.)
We want now to evaluate the evolution in (conformal) time of the
\textit{flavour vacuum} expectation value of the stress-energy tensor. Noticing that
the \textit{flavour vacuum} has been defined in (\ref{fv2}) by means of the operators $\hat{A}^{(a,b)}(\vp)$, whereas the fields $\hat{\psi}_i$ are defined at any time in terms of the operators $\hat{a}^{(a,b)}(\vp,\e)$, we have to find explicit relations between these two sets of operators that hold at any time $\e$.
This can be achieved by noticing~\cite{parker} that
\be
\hat{a}^{(a,b)}(\vp,\e)=\sum_{a=\pm c}D^{a}_{c}(p,\e)\hat{A}^{(c,abc)}(ac\vp)
\ee
with $D_{a}^{a'}(p,\e)$ defined through the equation
\bea\label{D}
	 D_{(a')}^{(a)}(p,\e)=\delta_{a'}^{a}+a'\int^{\e}_{\e_0}d\e'\frac{1}{4}\frac{\C'(\e')}{\sqrt{\C(\e')}}\frac{m p}{\w(\e')^2}\, e^{2 i a' \int\w(\e') 										
				d\e'}D_{(-a')}^{(a)}(p,\e')
\eea
with $a,a'=-1,1$.
Hence, at any time $\e$, the solution of the equation of motion (\ref{psiamin}) can be written as
\be\label{psi}
\hat{\psi}(\eta,\vec{x})=\left(\frac{1}{L \sqrt{\\C(\eta)}}\right)^{\frac{3}{2}} \!\!\!\!\!\!\!\!
\sum_{\mbox{\tiny{$\begin{array}{c} \vp  \\ a,b,c=\pm 1 \end{array}$}}}  \!\!\!\!\!\!\!\!
\hat{A}^{(a,b)}(\vp)D^{a}_{c}(p,\e) v^{(c,abc)}(ac\vp,\e) e^{ia\vp\cdot\vx-ic\!\int\!\!\w(\e) d\e}.
\ee
The functions $D^{a}_{b}(p,\e)$ encode the features of curved space-time, but an explicit solution of
equation (\ref{D}) for a generic conformal scale factor $\C(\e)$ is not known. In the rest of the work we will keep them
in their implicit form, with the understanding that explicit formulae have to be evaluated case by case. In particular, in our case of D-particle foam, in which for late eras of the Universe the conformal scale factor of the induced metric (\ref{confmetrtime}) is close to one, for small variances $\ll \sigma^2 \gg \, \ll \, 1$ of the D-particles, we can evaluate approximately such functions, in an appropriate expansion in powers of $\ll \sigma^2 \gg$.

These tools are sufficient to show that, in the continuous limit ($L\rightarrow \infty$), we have
\be\label{tiigen}
\lfv T_{ii}(\hat{\psi}_1,\hat{\psi}_2)\rfv =\lmv T_{ii}(\hat{\psi}_1,\hat{\psi}_2) \rmv+
\sst \!\! \int^{k_{\rm max}}_0 \!\!\!d p\left[V^2(p)\left(\mathcal{T}_{ii}(\e,p,m_1)+\mathcal{T}_{ii}(\e,p,m_2)\right)\right] +
\mathcal{O}(\ssst)
\ee
and
\be\label{t00gen}
\lfv T_{00}(\hat{\psi}_1,\hat{\psi}_2)\rfv =\lmv T_{00}(\hat{\psi}_1,\hat{\psi}_2) \rmv +\sst \!\! \int^{k_{\rm max}}_0 \!\!\! d p\left[V^2(p)\left(\mathcal{T}_{00}(\e,p,m_1)+\mathcal{T}_{00}(\e,p,m_2)\right)\right]+ \mathcal{O}(\ssst)
\ee
with all the other components vanishing. Above we used the notation
\bea \label{tcal}
	\mathcal{T}_{ii}(\e,p,m)&\equiv& \frac{1}{3}\frac{8}{(2 \pi)^2} \frac{p^4\sqrt{\Ce}}{\we}\left(\left(1-|D^{-1}_{1}|^2-|D^{1}_{-1}|^2\right)+
				\frac{2 m\sqrt{\C}}{p} Re\left[e^{-2i\intw} D^{-1}_{1}D^{-1*}_{-1}\right]\right) \nonumber \\
	\mathcal{T}_{00}(\e,p,m)&\equiv& \frac{8}{(2 \pi)^2} p^2  \we\sqrt{\Ce}\left(1-|D^{-1}_{1}|^2-|D^{1}_{-1}|^2\right) \nonumber \\
				\eea
and
\be
V^2(p)=\frac{\sqrt{p^2+m_1^2}\sqrt{p^2+m_2^2}-p^2-m_1 m_2}{\sqrt{p^2+m_1^2}\sqrt{p^2+m_2^2}}~.
\ee
As in the boson case~\cite{mavrosarkar}, we have also introduced a cutoff $k_{\rm max}$ in the four-momenta, since all expressions are affected by ultraviolet divergences.

Before discussing these divergences let us stress that in case of $\theta = 0$ and/or $m_1=m_2$ (\emph{i.e.} in the absence of mixing), the function $V^2(p)$ vanishes and therefore
$\lfv T_{\mu\nu}(\hat{\psi}_1,\hat{\psi}_2)\rfv =\lmv T_{\mu\nu}(\hat{\psi}_1,\hat{\psi}_2) \rmv$, consistently with what we would expect if we set $\hat{\psi}_1=\hat{\psi}_2$ in (\ref{fv}) and (\ref{G})~\cite{vitiello}.
It is important to remark once more that, in the context of the Heisenberg picture we have adopted here,
the flavour vacuum condensate $V(p)^2$, which encodes the structure of the flavour vacuum in terms of
Fock states in the mass representation~\cite{vitiello}, has been computed in the asymptotic past ($\e \to -\infty$), when the induced space-time (\ref{confmetrtime}) is assumed \emph{flat} (assuming a uniform D-particle background at $\e \to -\infty$, with no significant curvature effects. It is understood that such assumptions are highly model dependent, and hence the considerations here are specific to the initial conditions). In our Heisenberg picture, the term $V(p)^2$ does not evolve with time, and as such, its contribution will not be involved in any normal ordering.

\section{Normal Ordering \label{sec6} }

Let us now focus on the ultraviolet divergences.
These infinities, in local field theories, are conventionally removed by renormalization. In flat space-time this has been achieved by a suitable normal ordering. In the case of a conventional local field theory in a time dependent
metric background, the renormalization~\cite{birrell} would involve state dependent counter terms which, in a covariant
procedure, can be tensorially constructed from the metric tensor.
As already mentioned, in the present field theory model, which is the low-energy limit of a string theory involving D-particle capture of stringy matter~\cite{mavrosarkar} (c.f. fig.~\ref{fig:recoil}), the procedure of normal ordering is dictated by the underlying microscopic physics; the subtraction procedure, in particular,  has to be such that in the limit of the absence of D-particles and their fluctuations, $\ll \sigma^2 \gg \to 0$ the mixing phenomenon should disappear.
Moreover in our D-particle foam model one needs to distinguish two effects, as far as the structure of the underlying space-time is concerned. The first effect concerns a \emph{background} space-time, over which propagation of low-energy matter excitations (fermions or bosons) takes place. The background space-time in our case of D-particle foam has been argued to be obtained from quantum fluctuations of individual D-particles, which in the case of first quantised string framework are due to a summation over world-sheet topologies~\cite{szabo}, upon (statistically) averaging over populations of D-particles on the D3-brane world (c.f. fig.~\ref{fig:recoil}). This leads to a \emph{background} space-time metric of the form (\ref{confmetrtime}).

The individual MSW interactions of the flavoured matter excitations with the D-foam background~\cite{barenboim}, produce extra back-reaction local fluctuations on the space-time structure. They do not cause metric distortions, as already mentioned, but affect the particle mode's energy-momentum dispersion relations. This parallels and is in a similar spirit to the standard result when one considers particle production at the end of inflation~\cite{chung}. Hence in the effective QFT we have to take into account interactions of the neutrinos and the D-particle medium. In this respect, the scale factor $\C(\eta)$ appearing in the above formulae for the fermions, should be considered as representing the background space-time. The energies $\omega$ on the other hand will contain an ``\emph{effective}'' scale factor
\begin{equation}
\C_{\rm eff} (\eta) = \C(\eta) + \Delta \C(\eta) \equiv \C(\eta) \left(1 + \frac{\Delta \C(\eta)}{\C(\eta)}\right).
\label{effscale}
\end{equation}
In the dispersion relation
\be\label{oeff}
\omega_{\rm eff} = \sqrt{p^2 + m^2\C_{\rm eff}(\eta)} \simeq \sqrt{p^2 + m^2\C(\eta)} +
\frac{m^2\,\Delta \C(\eta)}{2\sqrt{p^2 + m^2\C(\eta)}}~,
\ee
to leading order in the approximation  $|\Delta \C| \ll 1 $. The $\Delta \C$ is the MSW contribution.
For the MSW scenario (c.f. (\ref{gravmsw2}), (\ref{u2msw})), one has the estimate
\be\label{dmfoam}
m^2 \Delta \C(\eta) \sim \C(\eta) \Delta m^2_{\rm foam} \simeq \C(\eta)\,{\cal N}^\star \, M_P \, p \sim \C(\eta)\,\ll \sigma^2 \gg g_s^{-1}\, M_s \,p ~.
\label{dc}
\ee
Such dispersion relations, that take proper account of the non-trivial interactions of the matter probes with the D-particles in the foam, should replace the free-particle dispersion relations considered in \cite{parker}
and used so far.

Notice that in our model, in general $\Delta \C$ would depend on both momenta and position of the mode.
For the purposes of this work, we are only interested in an order of magnitude estimate
of the induced dark energy contributions. We do not  need to specify the precise form of the induced distortion $\Delta \C$, apart from the fact that
in the context of our model we know that this will be proportional to the stochastic fluctuations of the recoil velocity, $\sigma^2(\eta)$, at the time of the interaction.
For sufficiently small $\ll \sigma^2 \gg$ one may ignore the momentum dependence of $\Delta \C(\eta)$ in (\ref{dc}).
On this basis, one can replace $\omega$ in Eqs.~(\ref{tcal}) by $\omega_{\rm eff}$ (\ref{oeff}), leaving the initial-time condensate $V(p)^2$ intact.
In a similar vein, eq. (\ref{D}), becomes:
\bea\label{Deffectivemass}      D_{(a')}^{(a)}(p,\e)&=&\delta_{a'}^{a}+a'\!\!\int^{\e}_{\e_0}\!\!\!d\e'      \Big(\sqrt{\C(\e')}\frac{p}{2 m_{\rm eff}(\e')}\frac{\w'(\e')}{\w(\e')}\, e^{2 i a' \!\!\!\int\!\!\!\,\w(\e')d\e'}D_{(\!-a'\!)}^{(a)}(p,\e')\Big)
\eea
with $a,a'=-1,1$ and $\w(\e)=\sqrt{p^2+m^2_{\rm eff}(\e)\C(\e)}$, and $m^2_{\rm eff} \C(\eta) \equiv m^2 \C_{\rm eff}(\eta)$.

After these necessary preliminaries, we are now in a position to discuss our subtraction (normal ordering) procedure.
The latter, will be \emph{defined} in such a way that in the \emph{absence} of any MSW interaction of the fermion matter with D-particles, the stress tensor should \emph{vanish}.
Notice that the MSW interactions, which are proportional to the density of defects, are disentangled from the space-time background $\C(\eta)$, in the sense that they contribute to mass shifts but they do not induce space-time distortions.
However, in view of (\ref{u2msw}), in our picture, the induced mass shifts are also proportional to the stochastic
fluctuations of the recoil velocity of the D-particles, which affects the
space time background, c.f. (\ref{confmetrtime}). Hence, by subtracting the MSW-like terms from the stress tensor, no further subtraction would be necessary in order to ensure that in a Minkowski space time the stress tensor would vanish.
Replacing, then, $\omega$ by $\omega_{\rm eff}$ in Eqs.~(\ref{tcal}) one should use (\ref{oeff}) to leading order in a small-$\Delta \C(\eta)$ expansion to determine the energy density and pressure of the fermion fluid.

Since in our D-particle case, the change $\Delta \C$ is proportional to the variance $\sigma^2(\eta) \ll 1$ of the recoil-velocity fluctuations in our weak space-time foam background (c.f. (\ref{u2msw})), only leading order contributions proportional to $\sigma^2$ should be taken into account. We observe that the terms involving $D_a^{a'}$
operators become irrelevant to this leading order, and so the latter should be replaced by their flat-space-time counterpart, which vanish (c.f. (\ref{D}), (\ref{Deffectivemass})). As a result, Eqs.~(\ref{tcal}) become:
\bea\label{tcalnorm}
&& :\mathcal{T}_{ii}(\e,p,m):  \approx  -\frac{4\sqrt{\C(\eta)} p^4}{3(2 \pi)^2} \Delta \C(\eta)\sum_{i=1}^{2}\frac{m_i^2}{\left(p^2+m_i^2\C(\eta)\right)^{\frac{3}{2}}} \nonumber \\
&& :\mathcal{T}_{00}(\e,p,m): \approx  \frac{4\sqrt{\C(\eta)} p^2}{(2 \pi)^2}\Delta \C(\eta) \sum_{i=1}^{2}\frac{m_i^2 } {\left(p^2 + m_i^2\C(\eta)\right)^{\frac{1}{2}}}\nn
&&
\eea
where the approximate sign indicates leading orders in $\Delta \C$.
Since, to leading order in $\sigma^2$, it is expected, rather generically, that $\ll \Delta \C(\eta) \gg \, \propto \, \ll \sigma^2 (\eta) \gg$,
the background scale factors $\C(\eta)$ in the above relation can be replaced by
constants (\emph{i.e}. flat space-time). Moreover, upon taking the statistical average $\ll \dots \gg$ of (\ref{tcalnorm}) over D-particle populations, to a good approximation for the weak space-time foam situations of interest, any momentum dependence of $\ll \Delta \C(\eta)\gg$ disappears; hence the latter quantity can be taken out of the momentum integrals in (\ref{tiigen}), (\ref{t00gen}). This will be understood in what follows.

On recalling that for a relativistic fluid $T_{00}/\C(\e)$ represents the \textit{energy density}, and $T_{ii}/\C(\e)$ the \textit{pressure}, we can easily see from (\ref{tcalnorm}), (\ref{tiigen}), (\ref{t00gen}) that our fermionic vacuum condensate behaves as a fluid with negative pressure, and positive energy density, with an equation of state that satisfies $-1/3<w<0$. This is the result of the opposite powers of $\omega$ (and hence $\omega_{\rm eff}$, according to our discussion above) appearing in a specific way in the pressure and energy expressions (\ref{tiigen}),(\ref{t00gen}), (\ref{tcal}).

\section{Equation of State, Vacuum Energy Estimates and Dynamical Momentum Cutoff \label{sec7}}

To determine the precise value of both the energy density and pressure, and hence the equation of state, it is necessary to have knowledge on the momentum ultraviolet cutoff $k_{\rm max}$ used to regulate ultraviolet infinities in flat space-time. To leading order in the small expansion $\Delta \C(\eta)$, such flat space-time approximation for the evaluation of the cutoff function proves sufficient. As discussed in \cite{mavrosarkar}, a \emph{dynamical} cutoff function appears if one considers \emph{particle production} due to the flavour vacuum.
In flat space-times, the particle number is given by:
\be
\lfv \hat{N}_{e}(\vp) \rfv=\sst \; \Xi(p)\left(1-\cos\left[ \left(\w_1+\w_2\right)\eta\right]\right)+\mathcal{O}(\ssst)+\mathcal{O}(\sigma^2)
\ee
with
\be
\Xi(p)=\frac{\left((\w_2- m_2)(w_1+ m_1)- p^2\right)^2}{2 \w_1 \w_2 (\w_2- m_2)(w_1+ m_1)}
\ee
and $\w_i=\sqrt{p^2+m_i^2}$, $i=1,2$.\begin{figure}[h]
\includegraphics[width=6cm]{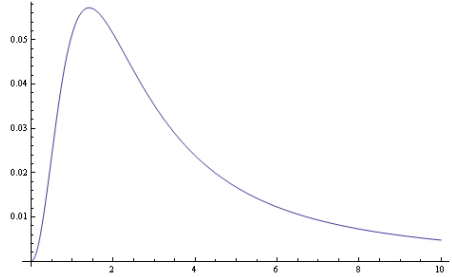}
\caption{The function $\Xi(p)$ is plotted in the range $p\in[0,10]$ for the values $m_1=1$ and $m_2=2$ (in arbitrary units, just for illustration purposes). Notice that there is a maximum~\cite{vitiello}, corresponding to the point $\sqrt{m_1 m_2}=\sqrt{2}$, in our arbitrary units.}
\label{fig:cutoff}
\end{figure}
The $p$ dependence is essentially determined by the behaviour of $\Xi(p)$. For large $p$ (compared to masses) the function falls off
with an inverse forth power of momentum (for a plot of this function vs. momentum, $p$, see fig.~\ref{fig:cutoff}):
\bea
\Xi(p) \approx  \frac{1}{2}\frac{(m_1^2-m_2^2)^2}{p^2} \, - \,
\frac{(m_1-m_2)^2((m_1+m_2)^2+2(m_1^2+m_2^2))}{8 p^4}
\eea
and so there is a scale (determined by the ratio of the two terms in the above)
\be\label{scale}
k_0 \approx\frac{1}{2}\sqrt{3 m_1^2+2 m_1 m_2+3 m_2^2}
\ee
which is a plausible cut-off scale in momenta $p$. Thus although there is no sharp cutoff, nevertheless, the flavour vacuum is populated significantly by fermionic models below this scale, and hence the latter serves as our cutoff $k_{\rm max}$, appearing in (\ref{tiigen}), (\ref{t00gen}). A similar situation characterised the ((1+1)-dimensional) boson case in \cite{mavrosarkar}. The reader is invited to compare the order of magnitude of (\ref{scale}) with
that of (\ref{bosons}).

It is easy to show that
{\small \bea\label{vacener}
\lfv: T_{00}(\hat{\psi}_1,\hat{\psi}_2):\rfv \approx \sst \frac{{\overline m}^{2}\,({\delta}m)^{2}}{\pi^{2}} \Delta \C(\eta) \textsl{I}(k_{\rm max})
\eea}where
\begin{equation}
    \textsl{I}(k_{\rm max})={\int_{0}}^{k_{\rm max}}dp \frac{p^{4}}{(p^{2}+ {\overline m}^{2})^{\frac{5}{2}}}
\end{equation}
and we have considered $\delta m \ll \overline{m}$.

Similarly, $\lfv:~T_{ii}(\hat{\psi}_1,\hat{\psi}_2):~\rfv$ is given by
{\small \bea
\lfv: T_{ii}(\hat{\psi}_1,\hat{\psi}_2):\rfv \approx -\sst \frac{{\overline m}^{2}\,({\delta}m)^{2}}{3 \pi^{2}} \Delta \C(\eta) \textsl{J}(k_{\rm max})
\eea}where
\begin{equation}
    \textsl{J}(k_{\rm max})={\int_{0}}^{k_{\rm max}}dp \frac{p^{6}}{(p^{2} + {\overline m}^{2})^{\frac{7}{2}}}.
\end{equation} Representing $k_{\rm max}$ in units of the characteristic neutrino mass scale $\overline{m}$, \emph{i.e.}
\begin{equation}
k_{\rm max} \equiv \kappa \, \overline{m}~, \quad \kappa > 0~,
\end{equation}
it is easy to show that
{\small \begin{eqnarray}\label{engpresinteg}
     \textsl{I}(\kappa \overline{m})=\log(\kappa +\sqrt{1+\kappa^{2}})-\frac{\kappa (3+4 \kappa^{2})}{3 (1+\kappa^{2})^{\frac{3}{2}}} \nonumber \\
     \textsl{J}(\kappa \overline{m})=\log(\kappa +\sqrt{1+\kappa^{2}})-\frac{\kappa (15+35 \kappa^{2}+ 23 \kappa^{4})}{15 (1+\kappa^{2})^{\frac{5}{2}}}~.
\end{eqnarray}}

Because of (\ref{dc}), we may write
(\ref{vacener}) as:
{\small \bea\label{vacener2}
&~&\lfv: T_{00}(\hat{\psi}_1,\hat{\psi}_2):\rfv \approx \sst \frac{({\delta}m^{2})^2}{\pi^{2}} \frac{\Delta m_{\rm foam}^2}{\overline{m}^2} \C(\eta) \textsl{I}(k_{\rm max})~.\nonumber \\ &~&
\eea}
The reader is invited to compare this expression with the corresponding one for the boson case, (\ref{bosons2}), upon taking (\ref{dmfoam}) into account.
Upon making the simplifying assumption that the foam is responsible for the whole of the experimentally observed mass differences of light neutrino species, we observe that the factor multiplying ${\rm sin}^2\theta \textsl{I}(k_{\rm max})$ is of order $\frac{(\delta m^2)^3}{\overline{m}^2}\C(\e)$. For late eras,  $\C(\e) \sim 1$ (in units of the present epoch scale factor of the Universe). For the biggest of the mass differences observed today~\cite{minos}, $\delta m_{23}^2 \sim 0.0027 ~{\rm eV}^2$ (in conventional notation), this factor is of order $7\times 10^{-118}~M_P^4$. Moreover the observed mixing ${\rm sin}^2\theta_{23}$ contributes factors slightly smaller than one (the current experimental data~\cite{minos} indicate ${\rm sin}^2(2\,\theta_{23}) \,>\, 0.87$ at 68\% confidence level), and, for the ranges of cutoffs considered above, the cutoff factors   $\textsl{I}(k_{\rm max})$ are of order ${\cal O}(10)$ at most. The accepted magnitude of the vacuum energy, claimed to have been observed today in the form of a positive cosmological constant, $\Lambda$, is
$\Lambda \sim 10^{-122}~M_P^4$ ($M_P \sim 10^{19}$ GeV). In order to reproduce such a value one needs
\[{\sin ^2}\theta \;\frac{{\Delta m_{foam}^2}}{{{{\overline m }^2}}} \sim 1.4 \times {10^{ - 4}}.\] This is compatible with other phenomenological tests of
space-time foam using neutrinos~\cite{barenboim}.  The reader is also invited to compare this result with the bound (\ref{cond2}), derived from stringy uncertainty considerations, in the case of a string mass scale $M_s = \mathcal{O}({\rm TeV})$,  upon taking proper account of the theoretical uncertainties
due to model dependence, as discussed there.

The equation of state of this fermionic fluid in the flavour vacuum is approximately determined by:
\begin{equation}\label{weqn}
w_F = - \frac{1}{3}\frac{\int_0^{k_{\rm max}} dp V^2(p) \sum_{i=1}^{2} \frac{m_i^2}{(p^2 + m_i^2)^{\frac{3}{2}}}}
{\int_0^{k_{\rm max}} dp V^2(p) \sum_{i=1}^{2} \frac{m_i^2}{(p^2 + m_i^2)^{\frac{1}{2}}}}.
\end{equation}
From (\ref{weqn}), to leading order in ${\delta}m$, we can deduce that $ w_F$ lies in the range
$ -1/3 < w_F < 0$ since
\begin{equation}\label{weqn2}
    w_F=-\frac{1}{3}\left(1-\frac{\kappa^{5}}{5 g(\kappa) (1+\kappa^{2})^{\frac{5}{2}}}\right)
\end{equation}where $g(\kappa)$ is a non-negative function given by
\begin{equation}
   g(\kappa)=\log(\kappa +\sqrt{1+\kappa^{2}}) -\frac{\kappa (3+4 \kappa^{2})}{3 (1+\kappa^{2})^{\frac{3}{2}}}.
\end{equation}The asymptotic approach of $w_F$ to $-\frac{1}{3}$ as $\kappa \rightarrow \infty $ is logarithmic with the cut-off and can be shown to be
\begin{equation}\label{weqn3}
    w_F=-\frac{1}{3}\left(1-\frac{1}{\log(\kappa)}\left\{\frac{1}{5}+\frac{7}{8\kappa^{4}}-\frac{1}{2 \kappa^{2}}\right\}\right).
\end{equation}For the cut off of (\ref{scale}), $w_F\simeq -.17$. As $\kappa$ is changed from $\sqrt{2}$
the value of $w_F$ rapidly asymptotes to $-\frac{1}{3}$, \emph{e.g.} with $\kappa=\sqrt{20}$, $w_F\simeq -.27$.

In $3+1$-dimensions, for the fermionic case,  the relative factor of 3 in the denominator of the
spatial components of the stress tensor (\ref{tiigen}) (as compared to the temporal component (\ref{t00gen})) is due exclusively to the spatial dimensionality. In the bosonic case, by contrast, the dimensionality of space is not relevant when evaluating the equation of state $w_B$ in terms of the appropriately normal ordered components of the corresponding stress tensor. Hence, in the latter case, we obtain $w_B \simeq -1$, for late eras.

\section{Discussion and Outlook \label{sec8}}

In this work
we have evaluated above the contributions from (3+1)-dimensional low-energy fermions to the flavour-Fock-space vacuum energy and pressure on a brane world punctured by D-particle defects. We have found that the pertinent liquid is not describing a cosmological constant vacuum.
The equation of state lies in the range $0 > w > -\frac{1}{3}$.
This does not lead to acceleration of the Universe, which requires $w < -\frac{1}{3}$.

Of course, in a D-brane setting, as that of fig.~\ref{fig:recoil} examined here,  there are many other contributions
to the D3-brane world, some of which are notably attributed to bulk D-particles~\cite{Dfoam2}. At late eras (relative to the time of the cosmically catastrophic brane collision corresponding to a Big Bang), there are also bulk contributions to the brane dark energy, which are due to
strings stretched between D-particle defects and the D3 brane world. The reader should recall~\cite{Dfoam2} that the contributions to the vacuum energy from the above processes are due to the perpendicular components of the relative velocity of the D-particle with respect to the D3-brane world in the bulk; the motion of a D-particle on the D3 brane world, parallel to its uncompactified components does not lead to any contribution. Hence it is plausible that these contributions are subdominant since in late epochs the motion of the D3 brane in the bulk is extremely slow. It is in this sense that the dominant contributions to the D3-brane vacuum energy could come from the above-described processes of capturing and splitting of flavoured string states on the brane, corresponding to the flavour Fock-space vacuum contributions evaluated above.

To ensure that the flavour vacuum contributions to the dark energy leads to \emph{accelerating} Universes at late epochs, as the current phenomenology indicates, one should have bosons simultaneously present with fermions.
In fact the bosonic and fermionic contributions to the vacuum energy and pressure are algebraically additive.
In the case of D-particle foam, only electrically neutral particles interact non trivially with the D-particle defects.

In the context of supersymmetric low-energy field theories, such as those derived in the low-energy limit of superstrings, the relevant bosons may be the \emph{sneutrinos},  the supersymmetric partners of neutrinos, which have large masses due to target-space supersymmetry breaking. However, the relative mass differences between mass eigenstates may be assumed sufficiently small, since the mass differences are independent of supersymmetry, especially if, according to our D-particle foam model, they are quantum-gravitational in origin. Hence even if the partners have a much greater mass, due to supersymmetry breaking, we may assume that, among different flavours, the \emph{same } small mass differences that characterize the fermionic excitations also characterize the bosonic superpartner flavours. In this sense, one has contributions to the vacuum energy density and pressure from the bosons, which are of the same order as those of fermions.
In realistic supersymmetric models, the total equation of state may be complicated, as it depends on the various fluids that participate in the flavour vacuum structure. Nevertheless, it is possible to have an equation of state that guarantees a late-era acceleration of the universe.

However, technically the extrapolation of the above results to supersymmetric cases is not a trivial task.
A supersymmetric theory is by construction typically an interacting theory, while above the quantization procedure adopted was based on free excitations. We hope to come back to this important issue in a forthcoming publication.

\section*{Acknowledgements} The work of N.E.M. and S.S. is partially supported by the European Union through the Marie Curie Research and Training Network \emph{UniverseNet} (MRTN-2006-035863), while that of W.T. by a King's College London (UK) graduate scholarship.

\end{document}